# Interaction of Acoustic and Optical Phonons in Soft Bonded Cu-Se Framework of Large Unit Cell Minerals with Anionic Disorders

*Kewal Singh Rana[1], Raveena Gupta[2], Debattam Sarkar[3], Niraj Kumar Singh[1], Somnath Acharya[4], Satish Vitta[4], Chandan Bera[2], Kanishka Biswas[3] and Ajay Soni[1]**

[1]School of Physical Sciences, Indian Institute of Technology Mandi, Mandi 175075, Himachal Pradesh, India

[2]Institute of Nano Science and Technology, *Knowledge City, Sahibzada Ajit Singh Nagar, Punjab - 140306, India*

[3]New Chemistry Unit, School of Advanced Materials and International Centre of Materials Science, Jawaharlal Nehru Centre for Advanced Scientific Research (JNCASR), Jakkur, Bangalore 560064, India

[4]Department of Metallurgical Engineering and Materials Science, Indian Institute of Technology Bombay, Mumbai 400076, India

Corresponding Author Email: ajay@iitmandi.ac.in

Large unit cell copper-chalcogenide based minerals with high crystalline anharmonicity have a potential for thermoelectric applications owing to their inherent poor lattice thermal conductivity. Here, the softening of copper-selenium bonding and hence crystal framework plays an important role in superionic conduction and thermal conductivity. We have studied $Cu_{26}Nb_2Sn_6Se_{32}$, $Cu_{26}Nb_2Sn_6Se_{31.5}$ and $Cu_{26}Nb_2Sn_6Se_{30}Te_2$ minerals with a strategically tailored anionic disorder. These compounds have *p*-type degenerate behavior with carrier concentration ranging between (2.7 – 15.3) x $10^{20}$ $cm^{-3}$, at 300 K, high power factor ~ 500 μW $m^{-1}$ $K^{-2}$ and low lattice thermal conductivity, ~ 0.70 W $m^{-1}$ $K^{-1}$, at ~ 640 K. The existence of two low frequency Raman active optical modes associated with soft Cu and Se atoms, three localized Einstein modes in specific heat, suggest high scattering between acoustic and optical branches with very short phonon lifetime (~ 0.3 – 0.6 *ps*). The excess vibrational density of states at low energies with compressed and flat optical branches strongly hinders the heat transport in these crystalline mineral. Comparatively, $Cu_{26}Nb_2Sn_6Se_{30}Te_2$ is a promising thermoelectric material because of high crystalline anharmonicity and softening of Cu-Se framework due to heavier tellurium atom.

Recent times, minerals have studied extensively due to their intrinsic properties of ultralow $\kappa_l$. In this regard, it is crucial to demonstrate the fundamental understanding of the heat transport inside the complex crystal structured mineral having crystalline nature. Enhanced phonon scattering is an essential requirement for poor thermal conductivity which can give a better thermoelectric performance and thermal managements in the electronic device. Thermoelectricity is a phenomena where the direct interconversion of heat and electricity is based on the Seebeck and Peltier effects.[1] The performance of the *TE* material depends upon a dimensionless figure of merit, $ZT = \left( S^2 / \rho\kappa_{total} \right) T$, where $S$ represents Seebeck coefficient, $\rho$ is the electrical resistivity, $T$ is the absolute temperature and $\kappa_{total}$ represents total thermal conductivity. Here, $\kappa_{total}$ is a sum of electrical ($\kappa_e$) and lattice ($\kappa_l$) part of the thermal conductivity.[1] All the *TE* parameters, except $\kappa_l$, are coupled together through charge carrier concentrations. For better conversion efficiency, *TE* materials should have maximum *ZT* which can be achieved by two ways, (i) by enhancement of the power factor ($PF = \left( S^2 / \rho \right)$) or good electrical transport, and (ii) by minimisation of $\kappa_{total}$ or poor thermal transport.[2] For this, the charge and heat transport have to be decoupled in the materials. However, the *PF* enhancement can be achieved by optimisation of charge carriers;[3-4] quantum confinements;[5] band engineering [6] and low-energy carrier filtering,[7] while the $\kappa_l$ can be minimized through solid-solution, alloying,[8] nano-structuring,[9] existence of soft-phonon modes,[10] nano-scale defects,[3] atomic rattlers,[11] large unit cell,[12] superionic compounds,[13] complex crystal structures,[14] and high bond anharmonicity.[15-16]

The chalcogen family of compounds (S, Se and Te) have shown greatest impact in many state-of-the-art *TE* materials in the mid-temperature range (500 – 700 K).[14, 17] Various binary, ternary and quaternary copper chalcogenides and their derivatives with transition elements (Fe, Mn, Zn, Co, Nb, V, Ni), have emerged as good *TE* materials.[14] The minerals such as Colusites $Cu_{26}A_2B_6S_{32}$ (A = Nb, V, Mo, W, Cr, Ta; B = Ge, As, Sn),[18-19] Chalcopyrites Cu(Ga, In)$Te_2$,[20] Tetrahedrites $Cu_{12-x}M_x(Sb,As)_4S_{13}$,[21] Argyrodites $A^{m+}_{(12-n)/m}M^{n+}X_6^{-2}$, (A = Ag, Cu; M = Si, Sn, Ge, Ga; and X = S, Se, Te).[22] superionic $MCrX_2$, (M = Ag, Cu; X = S, Se),[23] Sulfide bornite $Cu_5FeS_4$,[24] Kuramite $Cu_3SnS_4$,[25] and Kesterite $Cu_2$(Cd, Zn)Sn(S, Se)$_4$[26-27] show good *TE* performance due to complexity in the crystal structure.[14] The large number of atoms per unit cell, large crystal anharmonicity, low frequency optical modes, excess vibrational density of states and poor sound velocity altogether result in inherently low $\kappa_l$ through enhanced phonon scattering mechanism.[10, 22]

Colusite mineral has a simple cubic crystal structure having large number of atoms per unit cell, large *PF* and low $\kappa_l$ due to structural complexity.[19] In Colusites, high *PF* can be tuned with strong hybridization between the transition (Cu-3*d*) and chalcogen (S-3*p*) orbitals.[18, 28] Here, the presence of unoccupied states above the Fermi level results in a *p*-type semiconductor, the electron-deficient character can also be counted by the formal valances of $Cu^+$ (Z = 29, $3d^{10}$), $Nb^{5+}$ (Z = 41, $3d^0$), $Sn^{4+}$ (Z= 50, $4d^{10}$) and $S^{2-}$ (Z = 16, $3p^6$) ions[29]. Most of the available literature on Colusites focusses on chemical tailoring at the cationic sites keeping sulfur as the anion. It is observed that sulfur sublimation can induce large atomic scale defects, anti-site defects and interstitial defects.[19] The above discussed factors result in strong phonon scattering leading to reduction of $\kappa_l$ and consequently high *TE* performance.[18-19, 30-31]

In this study, we investigated the *TE* properties of $Cu_{26}Nb_2Sn_6Se_{32}$ (CNS-$Se_{32}$), $Cu_{26}Nb_2Sn_6Se_{31.5}$ (CNS-$Se_{31.5}$), and $Cu_{26}Nb_2Sn_6Se_{30}Te_2$ (CNS-$Se_{30}Te_2$) minerals prepared by melt grown method. A significant enhancement in *PF* and reduction in $\kappa_l$ is observed from CNS-$Se_{32}$ to CNS-$Se_{30}Te_2$ at high temperature. The low values of $\kappa_l$ in CNS-$Se_{30}Te_2$ is due to large anharmonicity and high phonon scattering from point defects and anion site disorders. The excess phonon density of states results in a boson peak at low temperature as investigated by heat capacity analysis. Further, the local Einstein modes interact with the low energy optical modes, suggest multi-phonon scattering in the material system. The large *N* results in the shrinkage of first Brillouin zone, which leads to folded back of high frequency vibrational modes as compressed optical modes with significantly reduction in the velocities (flat bands) near to zone boundaries.[32] The experimental work is supported by first principle calculations on electronic band structure and phonon dispersion, which also confirms the involvements of Cu and Se atoms and associated phonon modes for low $\kappa_l$.

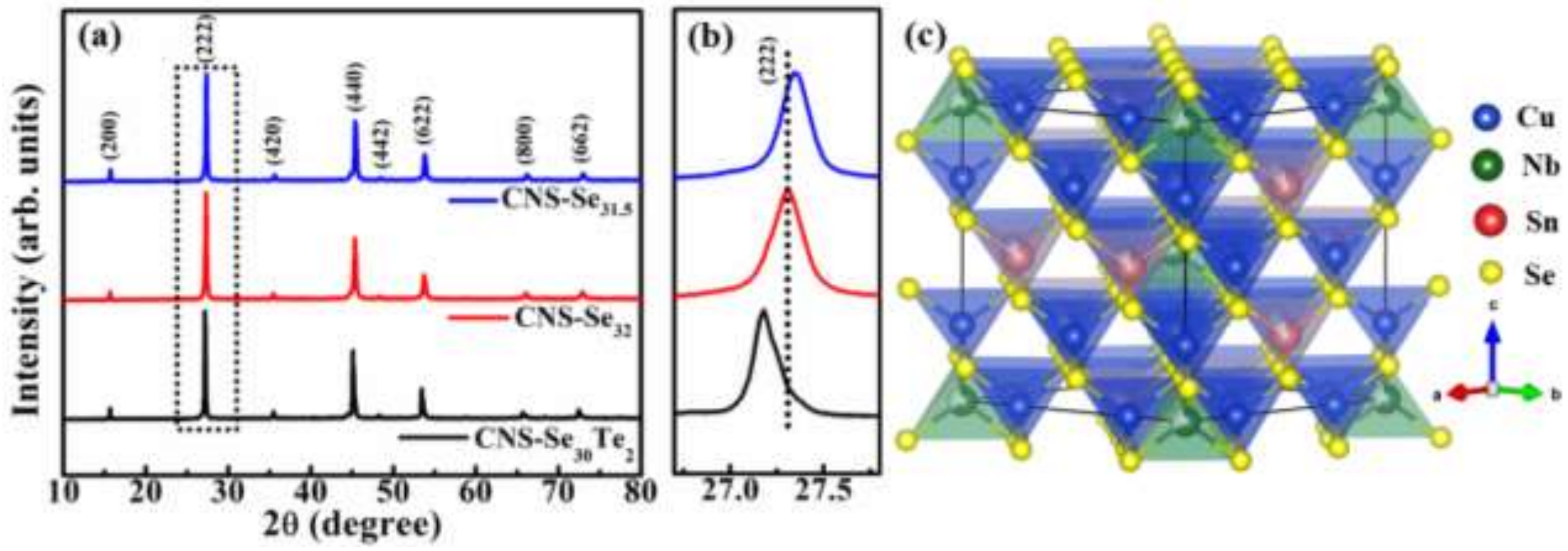

Figure 1. (a) Powder XRD pattern of CNS-$Se_{32}$, CNS-$Se_{31.5}$ and CNS-$Se_{30}Te_2$, (b) enlarged 222 peak showing lattice expansion of CNS-$Se_{30}Te_2$ and contraction of CNS-$Se_{31.5}$ with respect to CNS-$Se_{32}$, and (c) cubic crystal structure of CNS-$Se_{32}$.

XRD patterns of finely ground powders of hot pressed CNS-$Se_{32}$, CNS-$Se_{31.5}$ and CNS-$Se_{30}Te_2$ (**Figure 1 (a)**), show the crystalline nature and phase purity of the prepared samples.[33] All these compounds crystallize into cubic crystal structure with space group $P\bar{4}3n$. The Rietveld refinement of stoichiometric CNS-$Se_{32}$, (with goodness parameter, $\chi^2$ = 2.6), is shown in **Figure S1 (a)** and the estimated lattice parameters are ~ 11.3038 Å, ($a = b = c$) and unit cell volume ($V_c$) ~ 1444.35 $Å^3$. Further, the lattice parameters for Se deficient CNS-$Se_{31.5}$ are ~ 11. 2928 Å ($a = b = c$) and $V_c$ ~ 1440.14 $Å^3$, whereas for CNS-$Se_{30}Te_2$ are ~ 11. 3635 Å ($a = b = c$) with a largest $V_c$ ~ 1467.36 $Å^3$.

**Figure 1 (b)** shows the enlarged view of *222* peak (~ 27.3$^{o}$), where the peak shift with respect to CNS-$Se_{32}$ sample represents the contraction and expansion of lattice parameters due to introduction of Se vacancy in CNS-$Se_{31.5}$ and heavy Te atom in CNS-$Se_{30}Te_2$ sample.[18, 34] In comparison with the sulfur based Colusite, the lattice parameters and corresponding $V_c$ are increased due to involvement of larger radii Se and Te atoms.[28] The polyhedral cubic crystal structure of CNS-$Se_{32}$ with seven crystallographic sites is shown in **Figure 1 (c)**. Here, the Cu cations have three Wyckoff positions $Cu_I$ (12*f*), $Cu_{II}$ (8*e*) and $Cu_{III}$ (6*d*), while Nb and Sn cations are present at single site (2*a*) and (6*c*), respectively. Further, the Se anions have $Se_I$ (24*i*) and $Se_{II}$ (8*e*) Wyckoff positions.[35] The *x*, *y* and *z* positions of all the atoms with multiple Wyckoff positions are also shown in **Table S1**. All cations like Cu ($Cu_I$, $Cu_{II}$ and $Cu_{III}$), Nb and Sn are tetrahedrally coordinated with $Se_I$ and $Se_{II}$ anion and form a three dimensional (Cu/Nb/Sn)-$Se_4$ network.

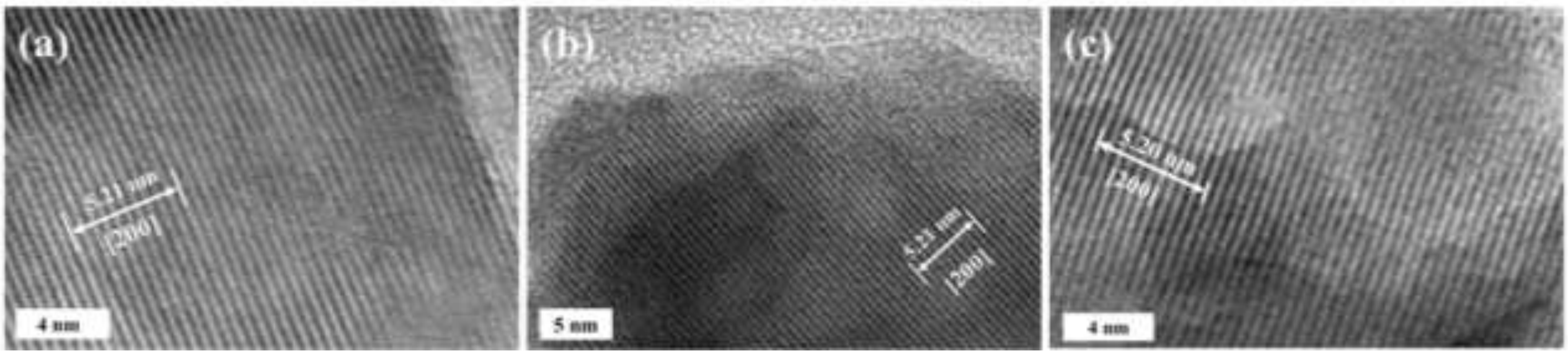

Figure 2. The bright field HR-TEM images of (a) CNS-$Se_{32}$; (b) CNS-$Se_{31.5}$ and (c) CNS-$Se_{30}Te_2$.

HR-TEM images in Figure 2 show that all three samples are crystalline in nature. While the XRD pattern shows the global crystalline nature, TEM images are the local representation of the the (*200*) planes of the crystal structures. Microstructural analysis and the elemental mapping of the constituent elements on the clean and polished surface, have been studied using the FE-SEM (**Figure S2 - S4**), which confirm the homogeneous distribution of elements, phase purity without any traces of segregation.

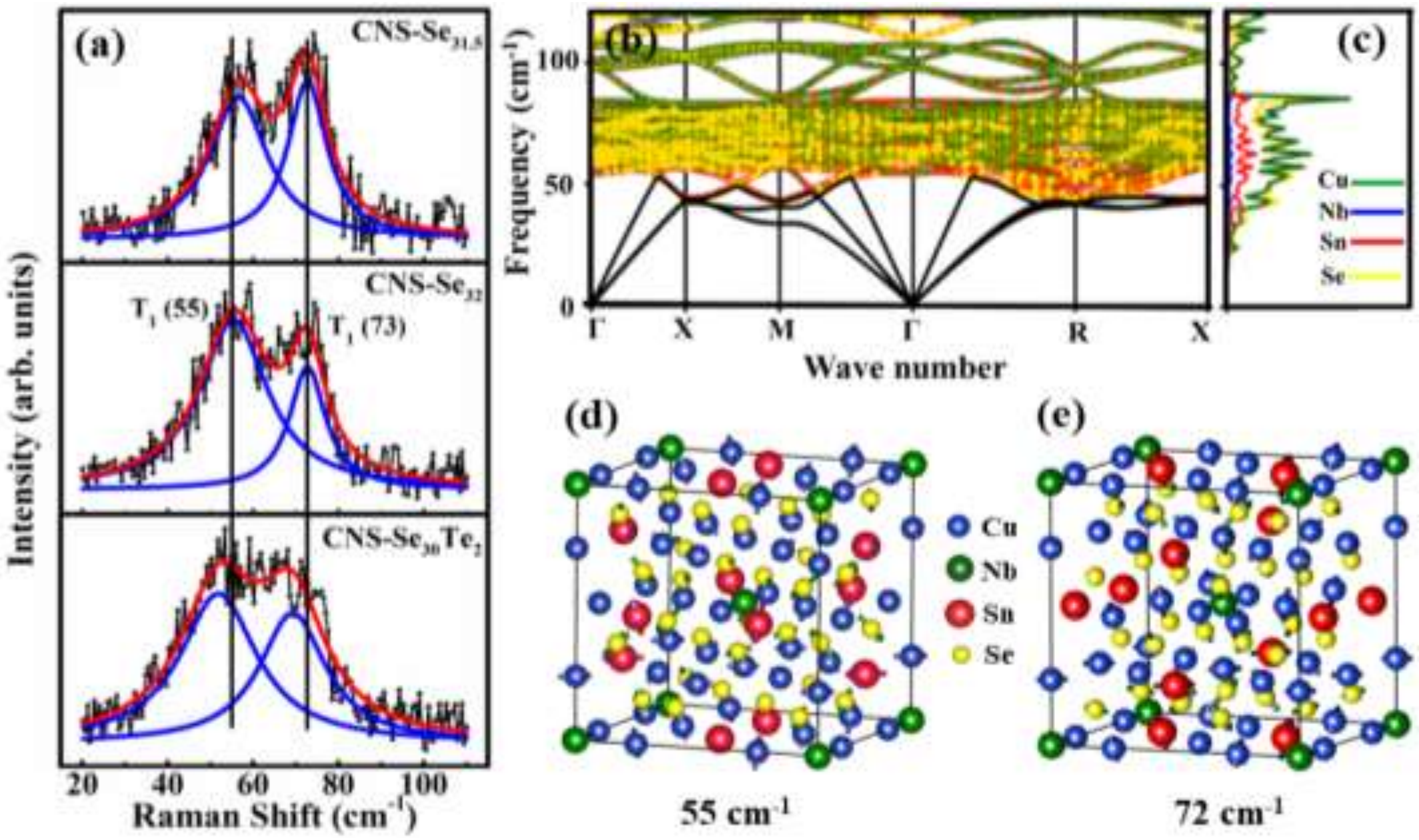

Figure 3. (a) Room temperature Raman spectra of CNS-$Se_{30}Te_2$, CNS-$Se_{32}$ and CNS-$Se_{31.5}$ in the low frequency region. The black curve represents the experimental data, blue and red curves are the individual and commutative peak fit. (b) the phonon dispersion curve, (c) the atom-projected *DOS* for CNS-$Se_{32}$, (d) and (e) the *Γ* point visualization of the low-lying optical modes of CNS-$Se_{32}$, respectively.

CNS-$Se_{32}$ has a cubic ($P\bar{4}3n$ space group) unit cell with 66 atoms (*N*) and thus have maximum (*3N*) ~ 198 phonon modes out of which (*3N-3*) ~ 195 are optical modes.[30, 36] From the group theoretical calculations, the Raman and IR active modes can possibly have $A_1$, $E$ and $T_2$ symmetries whose irreducible representation can be written as $\Gamma_{optical} = 5A_1 + 12E + 24T_2 + 24T_2$ (IR) where, $A_1$ are singly, $E$ are doubly and $T_2$ are triply degenerate modes.[19, 36-37] Here, we have observed two low frequency $T_1$ (~ 55 $cm^{-1}$) and $T_1$ (~ 73 $cm^{-1}$) Raman active modes for all the samples (**Figure 3 (a)**). The phonon dispersion and the atom-projected *DOS* calculations (**Figure 3 (b)** and **(c)**) shows the compressed, flat and several low frequency optical modes owing to large number of atoms in the unit cell. Furthermore, the strong acoustic-optical phonon interactions in the region of ~ 40 - 55 $cm^{-1}$ leads to more vibrational modes at *X*, *M* and *R* symmetry points. These low frequency (~ 40 - 80 $cm^{-1}$) flat modes are largely due to the mixed vibrations of Cu and Se atoms ($CuSe_4$ tetrahedra). The Cu and Se atoms has multiple Wyck off positions (describes as $Cu_I$, $Cu_{II}$, $Cu_{III}$, $Se_I$, $Se_{II}$) in the crystal lattice of CNS-$Se_{32}$ (**Figure S5**). At low frequency, the large variation of Cu-Se bond lengths with inhomogeneous bond strength in $CuSe_4$ tetrahedra soften the Cu-Se bonds (**Table S2**). **Figure 3 (d)** and **(e)** shows the visualization of both the low-lying modes of CNS-$Se_{32}$. In the CNS-$Se_{30}Te_2$, both modes shift toward the lower wave number (phonon softening), significantly, which is due to the change in the bond strength through the involvement of heavier Te atom and presence of strong crystal anharmonicity. To study the temperature response of Raman active modes, we have performed the temperature dependent Raman measurements (shown in **Figure S6**) and calculated the phonon lifetime ($\tau_i = \frac{1}{2\pi FWHM_i}$) for the modes from full width at half-maximum (*FWHM*). The estimated $\tau_i$ is found to be in the range of ~ 0.3 to 0.7 picosecond, ultrashort, which also remains constant in the temperature range from 300 K to 500 K (**Table S3**). The low $\tau_i$ further suggests smaller phonon mean free paths, faster scattering rate and ultimately hinder the heat carrying phonon transportation inside the crystalline solids.[38] Thus, being lightest element in the unit cell, Cu with the soft Cu-Se bonds is expected to affect the transport and vibrational properties, significantly.

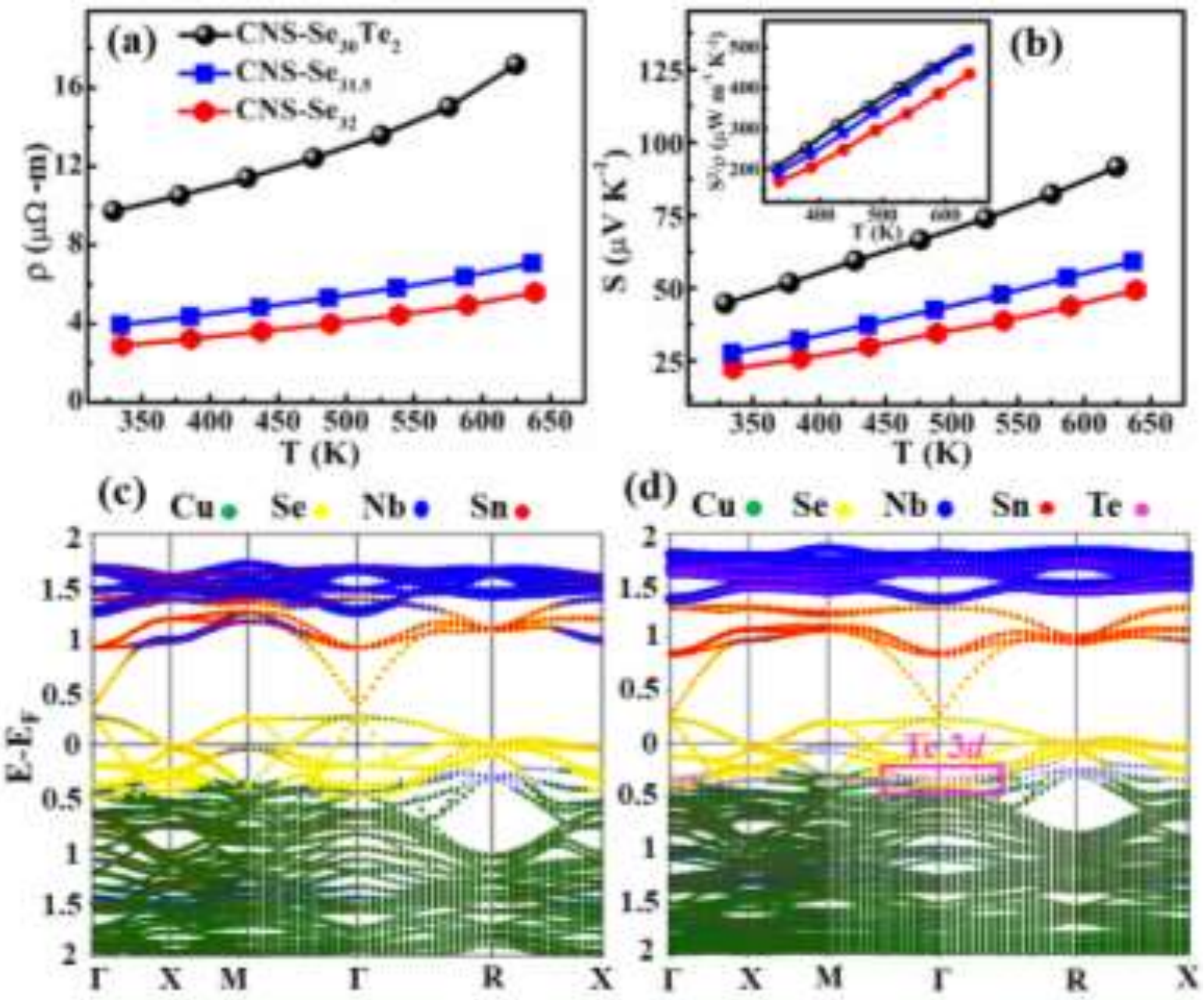

Figure 4. Temperature dependence of (a) ρ, (b) S, inset represents the power factor, $S^2/\rho$, and electronic band structure of (c) CNS-$Se_{32}$ and (d) CNS-$Se_{30}Te_2$.

Temperature dependent resistivity, $\rho$(T), increases with increasing temperature suggesting the degenerate semiconductor like nature of charge transport (**Figure 4 (a)**), which is comparable with the earlier studies of sulfur based Colusites.[39] Here, stoichiometric CNS-$Se_{32}$ has the lowest $\rho$(T) due to the ordered unit cell, whereas CNS-$Se_{30}Te_2$ has highest $\rho$(T) because of presence of Te anions in Se site. The room temperature Hall carrier density are ($n_H$) ~ 1.53, 1.08 and 0.27 × $10^{21}$ $cm^{-3}$ and mobility ($\mu_H$) are ~ 6.25, 5.37 and 11.80 $cm^2$ $V^{-1}$ $s^{-1}$ for CNS-$Se_{32}$, CNS-$Se_{31.5}$ and CNS-$Se_{30}Te_2$ respectively. Clearly, the Se vacancies are electron donor which are decreasing the $n_H$ with compensation while presence of Te creating more disorders in the samples. The positive *S* (T) shows that holes are the majority charge carriers and the CNS-$Se_{30}Te_2$ has the highest *S* at high temperature (**Figure 4 (b)**). Here, the $n_H$ is one order less for CNS-$Se_{30}Te_2$ as compared to CNS-$Se_{32}$, which justifies the high $\rho$(T) and *S*(T) in CNS-$Se_{30}Te_2$. The overall power factor ($S^2/\rho$) shows that CNS-$Se_{30}Te_2$ has the highest *PF*. Further, CNS-$Se_{31.5}$ has the highest *PF* ($S^2/\rho$) ~ 494 μW $m^{-1}$ $K^{-2}$ almost equivalent to CNS-$Se_{30}Te_2$ having ~ 488 μW $m^{-1}$ $K^{-2}$, whereas CNS-$Se_{32}$ have *PF* ~ 434 μW $m^{-1}$ $K^{-2}$ at high temperatures (inset of **Figure 4 (b)**). The metallic $\rho$(T) temperature dependence for all the samples is because of the identical electronic structure near the Fermi level ($E_F$). The unoccupied states above the $E_F$ leads to *p*-type behavior of the compounds. The contribution of various atoms toward the electronic density of states above and below the $E_F$ for CNS-$Se_{32}$ and CNS-$Se_{30}Te_2$ are shown by different colors in **Figure 4 (c)** and **(d)**. In CNS-$Se_{30}Te_2$, the contribution of *3d* states of Te towards the electronic transport results to change in the total density of states (**Figure S7**), which enhances the *S* (T).

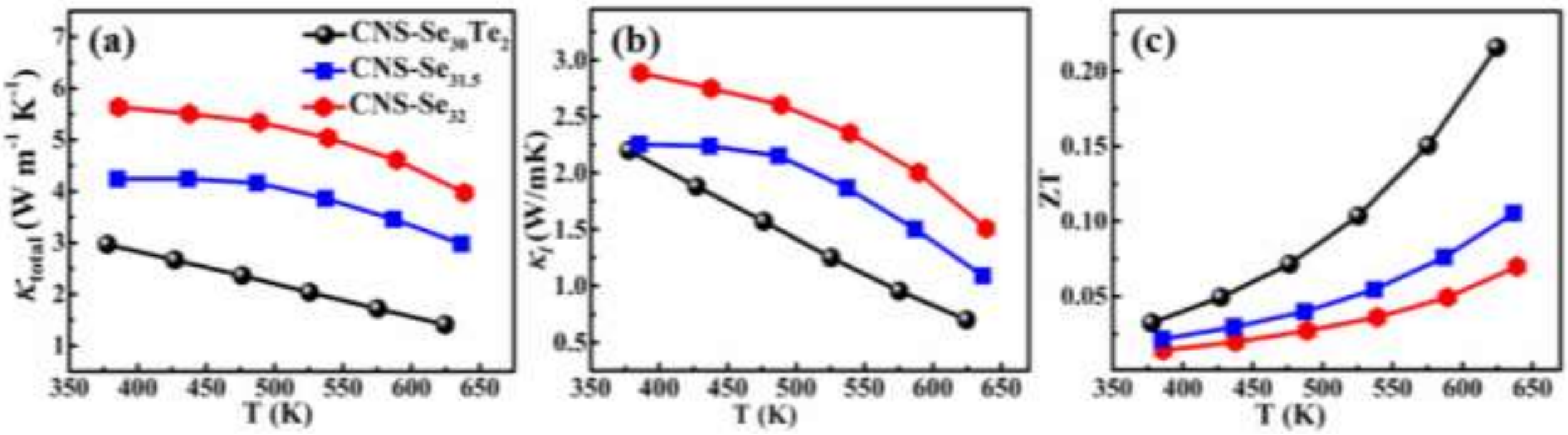

Figure 5. Temperature dependence of (a) $\kappa_{total}$, (b) $\kappa_l$ and (c) *ZT*.

The temperature dependence of $\kappa_{total}$ shows a decreasing trend with the rise in temperature (**Figure 5 (a)**). Here, the vacancies and the point defects at the anionic sites in CNS-$Se_{31.5}$ offers additional phonon scattering sites leading to reduction in $\kappa_{total}$, when compared with CNS-$Se_{32}$. On the other hand, for CNS-$Se_{30}Te_2$, the mass contrast as well as size difference offered by Se and Te atoms due to difference in the atomic radii (~ 190 pm (Se) and ~ 210 pm (Te)) enhances the phonon scattering leading to large reduction in the $\kappa_{total}$. The details of the thermal transport is presented in supplementary information **section S7**. Furthermore, the lattice contribution, $\kappa_l$ shows 1/T dependence due to anharmonic *Umklapp* processes (**Figure 5 (b)**).[40] At high temperature, CNS-$Se_{30}Te_2$ has the relatively lowest $\kappa_l$ ~ 0.70 W $m^{-1}$ $K^{-1}$ due to high disorder and overall *ZT* (~ 0.22), which is three times higher than the stoichiometric CNS-$Se_{32}$ at 625 K (**Figure 5 (c)**. To understand better the inherent poor $\kappa_{total,}$ we have analyzed the low temperature heat capacity data and extract the possible interactions of acoustic and optical phonons.

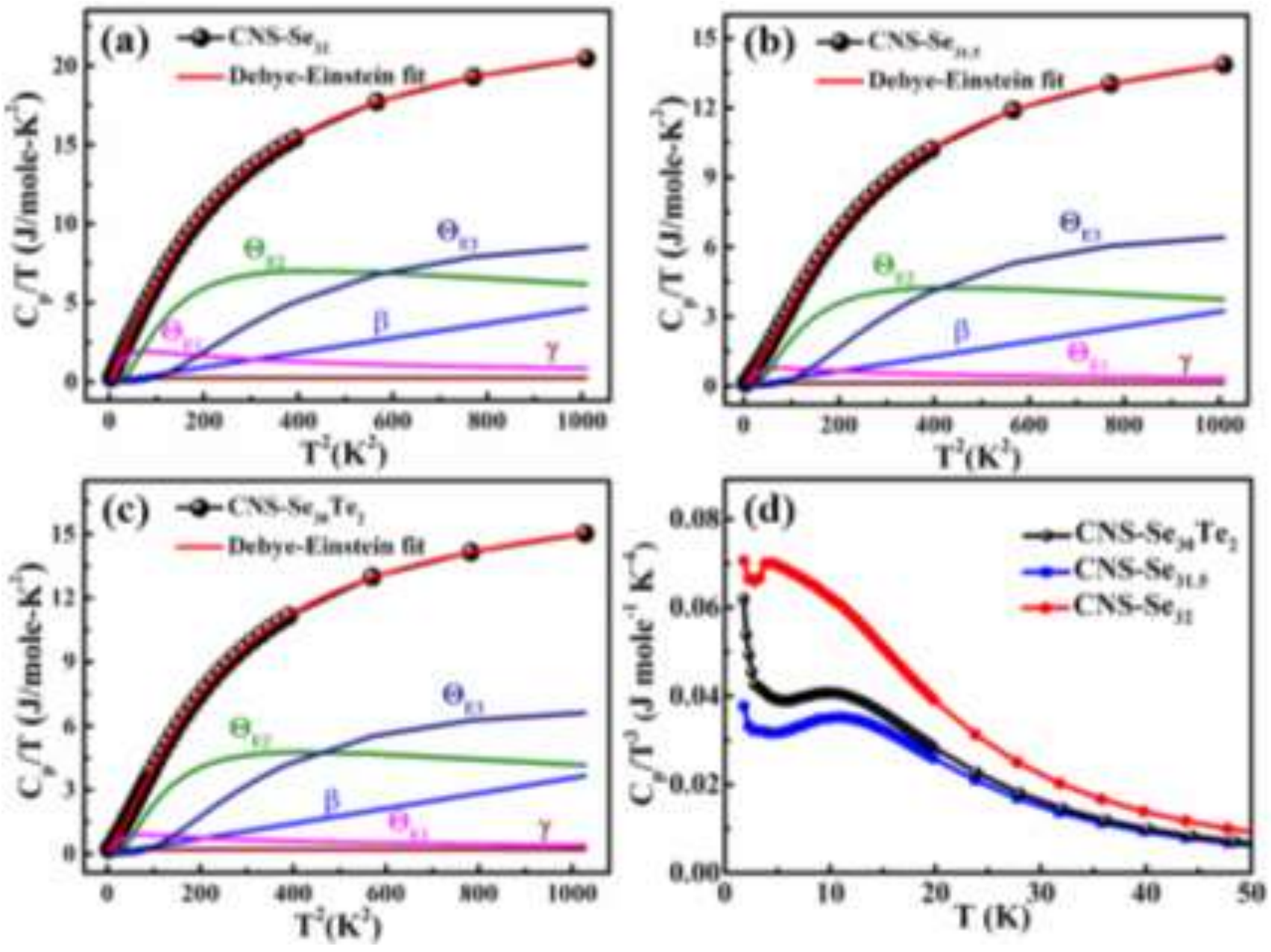

Figure 6. Temperature dependence of heat capacity fitted with one Debye and three Einstein modes for (a) CNS-$Se_{32}$, (b) CNS-$Se_{31.5}$, (c) CNS-$Se_{30}Te_2$ and (d) $C_p/T^3$ vs $T$.

The temperature dependence of $C_p$ at low temperature is shown in **Figures 6 (a) – (c)**. The anomalous negative value of electronic term ($\gamma$) in fitting with Debye model indicated that the $C_p$ analysis can't be examined from single Debye mode.[41] Therefore, we have fitted the data using one Debye-three Einstein (1*D*-3*E*) mode model using the following equation:

$$\frac{C_p}{T} = \gamma + \beta T^2 + A\Theta_{E1}^2 T^{2\left(-\frac{3}{2}\right)} \frac{e^{\frac{\Theta_{E1}}{T}}}{\left(e^{\frac{\Theta_{E1}}{T}}-1\right)^2} + B\Theta_{E2}^2 T^{2\left(-\frac{3}{2}\right)} \frac{e^{\frac{\Theta_{E2}}{T}}}{\left(e^{\frac{\Theta_{E2}}{T}}-1\right)^2} + C\Theta_{E3}^2 T^{2\left(-\frac{3}{2}\right)} \frac{e^{\frac{\Theta_{E3}}{T}}}{\left(e^{\frac{\Theta_{E3}}{T}}-1\right)^2}$$

where γ and β represents the electronic and lattice contribution of heat capacity; *A*, *B* and *C* are the material constants, and $\Theta_{E1}$, $\Theta_{E2}$ and $\Theta_{E3}$ are the three Einstein temperatures.[42] The obtained $\Theta_E$for CNS-$Se_{32}$ are 22.76 K (15.82 $cm^{-1}$), 53.68 K (37.31 $cm^{-1}$) and 97.92 (68.06 $cm^{-1}$) as shown in **Table S4**. The $\Theta_{E1}$, $\Theta_{E2}$ and $\Theta_{E3}$ individually contributes to the total $C_p/T$ where, $\Theta_{E1}$*'s* dominates below ~ 7 K, $\Theta_{E2}$*'s* from ~ 7 – 23 K and $\Theta_{E3}$*'s* above ~ 23 K. The energy of the Einstein mode is equivalent to the observed low frequency Raman active modes $T_1$ (~ 55 $cm^{-1}$) and $T_1$ (~ 73 $cm^{-1}$), and thus the interactions of these modes provide extra phonon scattering in these samples.

The obtained Debye temperature is shown in **section S8** and the estimated average sound velocity is found to be $\upsilon_{avg}$ ~ 2866, 3228 and 3165 m $sec^{-1}$ for CNS-$Se_{32}$, CNS-$Se_{31.5}$ and CNS-$Se_{30}Te_2$, respectively.[19]. The presence of localised Einstein modes, strong acoustic and optical phonon scattering, low $\upsilon_{avg}$, excess vibrational density of states at low temperature strongly affects the $\kappa_l$ of the materials. Thus, the interaction of low energy phonons results in poor thermal transport and $\kappa_l$. The excess *VDOS* at low temperature leads to the deviation in Debye $T^3$ law which is universal feature in the system having glass like thermal properties and the peak is called as "boson peak" (*BP*) as shown in **Figure 6 (d)**.[43-44] For this non-Debye behaviour the acoustic and phonon interaction plays a very crucial role which are associated with soft/loosely bonded Cu atoms.[43-46] From stoichiometric (CNS-$Se_{32}$) to non-stoichiometric compound (alloy CNS-$Se_{30}Te_2$ and deficient CNS-$Se_{31.5}$), the $C_p/T^3$ decreases remarkably, as the disorder and fragility increases due to alloying and deficient nature.[43]

Beside these factors the low $\kappa_l$ can also be understood by the crystal anharmonicity which is governed by the term Grüneisen parameter ($\gamma_G$) which relates the crystal volume and phonon frequency.[15] The large crystal anharmonicity in the compound causes large damping effects and enhances the scattering. We calculated the $\gamma_G$ from $\gamma_G = \left((3.1\,x\,10^{-6})\frac{M_a \delta \Theta_D^3}{\kappa_L N^{2/3} T}\right)^{1/2}$, where $M_a$ is the average atomic mass in amu, $\delta^3$ is the atomic volume (volume per unit), $\Theta_D$ the Debye temperature, $\gamma_G$ the Gruneisen parameter, $N$ the total number of atoms present in the primitive unit cell, and $T$ the temperature.[47] We obtained $\gamma_G$ ~ 1.16, 1.64 and 2.06, for CNS-$Se_{32}$, CNS-$Se_{31.5}$ and CNS-$Se_{30}Te_2$, respectively, and the values are higher than many state-of-the-art *TE* materials like $Bi_2Te_3$, $Sb_2Te_3$ PbTe, PbSe and PbS.[48] The significant enhanced $\gamma_G$ (~ 80%) in CNS-$Se_{30}Te_2$ than CNS-$Se_{32}$ is due to presence of disorder and mass difference at anionic site as both anions (Se and Te) are present at same site (24*i* and 8*e*) and due to the change in crystal volume.[49]

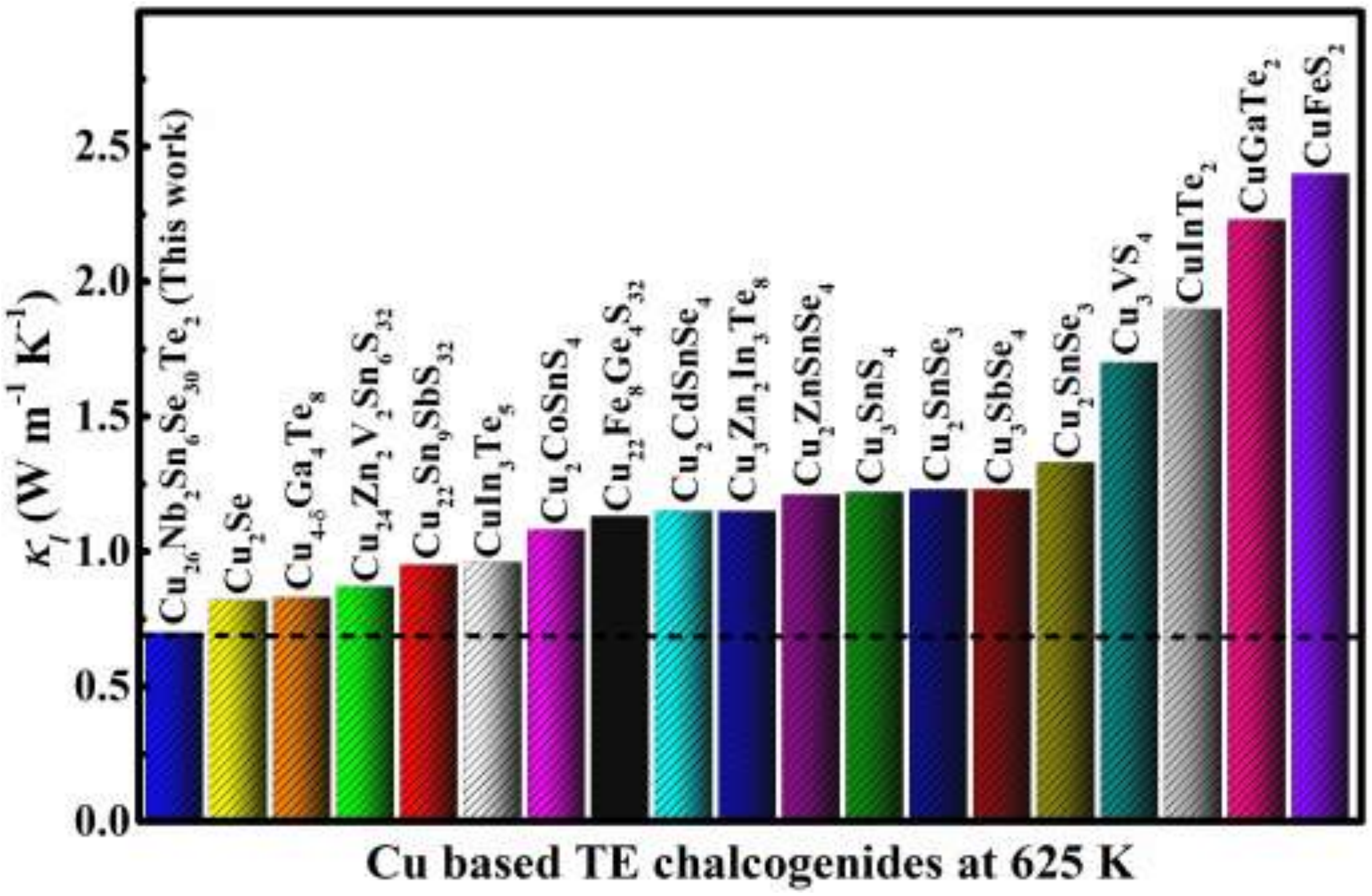

Figure 7. Comparison of the $\kappa_l$ of CNS-$Se_{30}Te_2$ sample with other copper-based chalcogenides at 625 K.[17-18, 20, 25, 50-61]

A comparative study of $\kappa_l$ of CNS-$Se_{30}Te_2$ with different Cu-based binary, ternary and quaternary chalcogenides is shown in **Figure 7**. The lower value of $\kappa_l$ can be understood due to relatively large

Grüneisen parameter, soft bonded Te atoms in Cu-Se crystal framework and softening of Raman active modes in CNS-$Se_{30}Te_2$ in comparison to CNS-$Se_{32}$, CNS-$Se_{31.5}$ and above shown compounds. Thus, the interaction of low energy acoustic and optical phonons results in poor thermal transport and $\kappa_l$.

In summary, the scattering of acoustic and optical phonons in Se based mineral is accountable for inherently poor $\kappa_l$. The Grüneisen parameter in CNS-$Se_{30}Te_2$ suggests strong crystal anharmonicity leading to $\kappa_l$ ~ 0.7 W $m^{-1}$ $K^{-1}$ at 625 K, where the structural disorder induced by the Te atom enhanced the phonon scattering. The excess vibrational density of state result to the appearance of Boson peak in the heat capacity data. The low-frequency optical modes are mainly associated with soft bonded Cu-Se tetrahedra. Additionally, the phonon lifetime ~ *ps*, confirms the high scattering rate causing a low thermal conductivity in these compounds. The *TE* performance of these materials can further be improved by enhancing the *PF* through band engineering and optimising the carrier concentrations.

## Experimental details

Polycrystalline samples of CNS-$Se_{32}$, CNS-$Se_{31.5}$ and CNS-$Se_{30}Te_2$ were synthesized via solid state melt-grown technique.[4, 13, 15] High purity (~ 99.99%) elemental Cu, Nb, Sn, Se and Te were weighed in stoichiometric ratio and placed in quartz tubes of ~ 10 mm diameter, and the tubes were flame-sealed at a vacuum of ~ $10^{-5}$ mbar. The sealed tubes were heated to ~ 1323 K with a rate of ~ 0.3 K $min^{-1}$ and kept at ~ 1323 K temperature for ~ 48 hours. Later, the tubes were cooled down to ~ 673 K with a rate of ~ 0.6 K $min^{-1}$, followed by further cooling to room temperature with a rate of ~ 1 K $min^{-1}$. The obtained ingots were ground and later subjected to hot-pressing at 723 K under a uniaxial pressure of 30 MPa in Ar atmosphere for a period of ~ 1 hour, to get the pellets for transport measurements. Mass density ($d_m$) of the obtained hot-pressed pellets were found to be ~ 5.70 g $cm^{-3}$ for CNS-$Se_{32}$ and CNS-$Se_{31.5}$; ~ 5.65 g $cm^{-3}$ for CNS-$Se_{30}Te_2$, as assessed via Archimedes principle. Crystal structure and phase purity study were identified by x-ray diffraction using a Rigaku Smart lab make rotating anode diffractometer through $CuK_\alpha$ as radiation source ($\lambda$ ~ 1.5406 $\AA$) with the scan rate of 2° $min^{-1}$ and the step size of 0.02°. The Hall measurements and the heat capacity ($C_p$) were performed using physical property measurement system (PPMS, Quantum Design). The high temperature, (320 K - 640 K), resistivity ($\rho$) and Seebeck coefficient ($S$) measurements were performed using ZEM-3 ULVAC instrument on bar-shaped pellets. Further, $\kappa_{total}$ was calculated using the relation, $\kappa_{total} = DC_pd_m$, (where $D$ and $C_p$ be the thermal diffusivity and specific heat respectively) estimated by NETZSCH-Laser flash apparatus under nitrogen atmosphere. Field emission scanning electron microscope and energy dispersive X-ray spectroscopy of the polished pellets were examined from JFEI, USA, make Nova Nano SEM-450. High resolution transmission electron microscope (HR-TEM) was performed through FEI Tecnai G2 20S-twin microscope operating at 200 kV under vacuum conditions. The Raman Spectroscopic measurements were carried out using Jobin-Yvon Horiba LabRAM HR evolution Raman spectrometer, where 532 nm excitation laser was used to excite the samples and Raman spectra were recorded with 1800 grooves/mm grating and ultra-low frequency filter was used for eliminating Rayleigh line to access low frequency Raman modes. Linkam stage was used for high temperature dependent (300 to 500 K) Raman studies. Furthermore, the details of the theoretical calculations are discussed in the **section S1** of supporting information.

**Author information ORCID:**

Kewal Singh Rana: 0000-0003-0345-6225

Niraj Kumar Singh: 0000-0003-2608-4022

Raveena Gupta: 0000-0002-7885-7026

Somnath Acharya: 0000-0003-0118-4326

Satish Vitta: 0000-0003-4138-0022

Kanishka Biswas: 0000-0001-9119-2455

Chandan Bera: 0000-0002-5226-4062

Ajay Soni: 0000-0002-8926-0225

**Author Information's:** The authors declare no competing financial interest.

**Acknowledgements:** A.S. acknowledges DST-SERB India (Grant No. CRG/2018/002197) for funding support and IIT Mandi for research facilities.

## Supporting Information

# Interaction of Acoustic and Optical Phonons in Soft Bonded Cu-Se Framework of Large Unit Cell Minerals with Anionic Disorders

*Kewal Singh Rana[1], Raveena Gupta[2], Debattam Sarkar[3], Niraj Kumar Singh[1], Somnath Acharya[4], Satish Vitta[4], Chandan Bera[2], Kanishka Biswas[3] and Ajay Soni[1]**

[1]School of Physical Sciences, Indian Institute of Technology Mandi, Mandi 175075, Himachal Pradesh, India

[2]Institute of Nano Science and Technology, *Knowledge City, Sahibzada Ajit Singh Nagar, Punjab - 140306, India*

[3]New Chemistry Unit, School of Advanced Materials and International Centre of Materials Science, Jawaharlal Nehru Centre for Advanced Scientific Research (JNCASR), Jakkur, Bangalore 560064, India

[4]Department of Metallurgical Engineering and Materials Science, Indian Institute of Technology Bombay, Mumbai 400076, India

Corresponding Author Email: ajay@iitmandi.ac.in

## Section S1: Computational Methods

The electronic structures were investigated using density functional theory (DFT) within the Vienna Ab initio Simulation Package (VASP) framework. [1-2] The projector-augmented-wave (PAW) method, [1, 3] combined with the Perdew-Burke-Ernzerhof (PBE)[4] approximation for the exchange and correlation terms of the functional, was employed. A plane-wave cut-off of 400 eV was set, and the structures were relaxed using a 4x4x1 Monkhorst-Pack grid for integrations over the Brillouin zone (BZ). A tolerance of $10^{-8}$ eV was established for the total energy and band energies, and calculations were continued until the absolute value of all components of the force was lower than 0.01 eV $Å^{-1}$. To determine the phonon band structure, a real-space approach was utilized to extract the force constants, as implemented in the Phonopy package.[5] For calculating the interatomic force constants, 2x2x1 supercells and displacements with an amplitude of 0.01 Å were used, and forces were computed with VASP.

## Section S2: XRD analysis

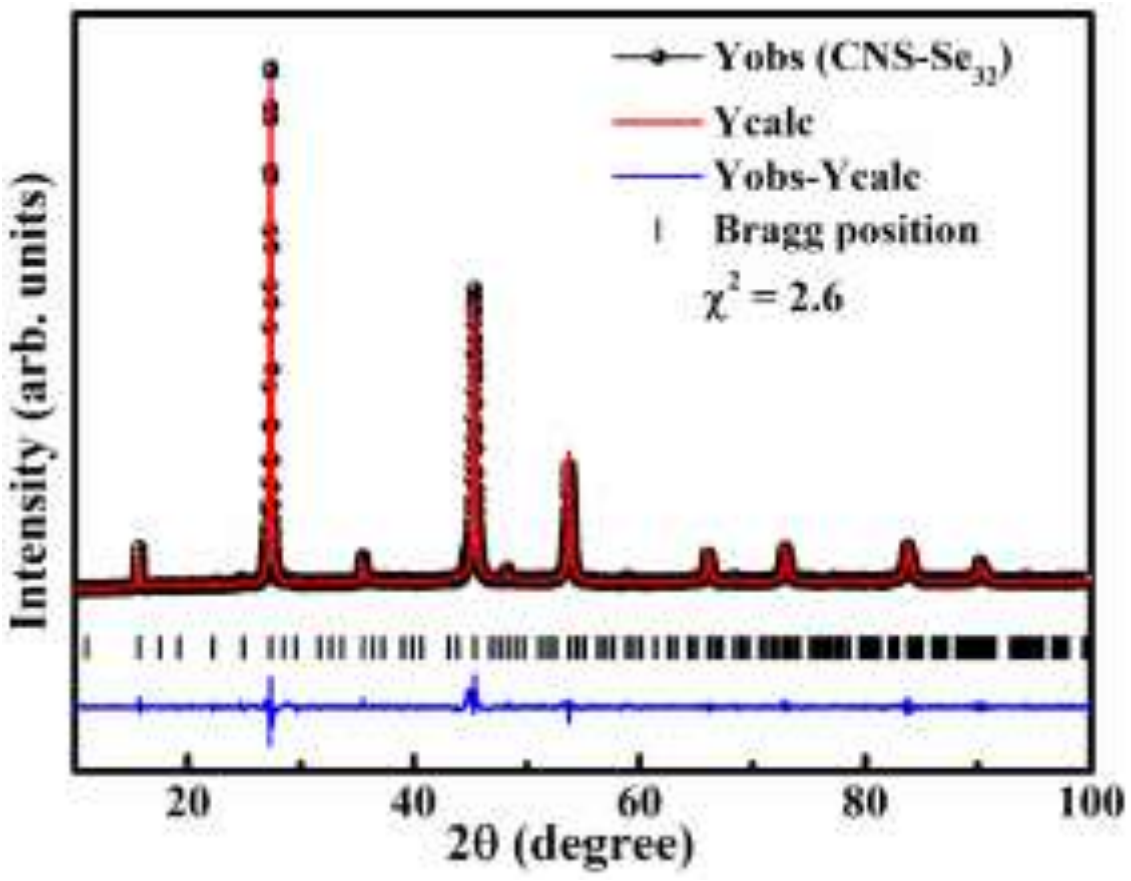

Figure S1. (a) Rietveld refined XRD pattern of CNS-$Se_{32}$, examined through by Full Prof Suite software.[6]

**Table S1.** Rietveld refined structural parameters of CNS-$Se_{32}$.

| **Atom** | ***x*** | ***y*** | ***z*** | **Wyckoff position** |
|---|---|---|---|---|
| CuI | 0.25488 | 0.00000 | 0.00000 | 12*f* |
| $Cu_{II}$ | 0.25023 | 0.25023 | 0.25023 | 8*e* |
| $Cu_{III}$ | 0.25000 | 0.00000 | 0.50000 | 6*d* |
| $Se_{I}$ | 0.12333 | 0.37465 | 0.37336 | 24*i* |
| $Se_{II}$ | 0.13062 | 0.13062 | 0.13062 | 8*e* |
| Sn | 0.25000 | 0.50000 | 0.00000 | 6*c* |
| Nb | 0.00000 | 0.00000 | 0.00000 | 2*a* |

**Section S3: FE-SEM analysis**

The FE-SEM analysis of CNS-$Se_{32}$, CNS-$Se_{31.5}$ and CNS-$Se_{30}Te_2$ samples are shown in **Figure S2 - S4** confirms that the distribution of all the elements are uniform throughout the sample as there are no poor or rich concentration of elements.

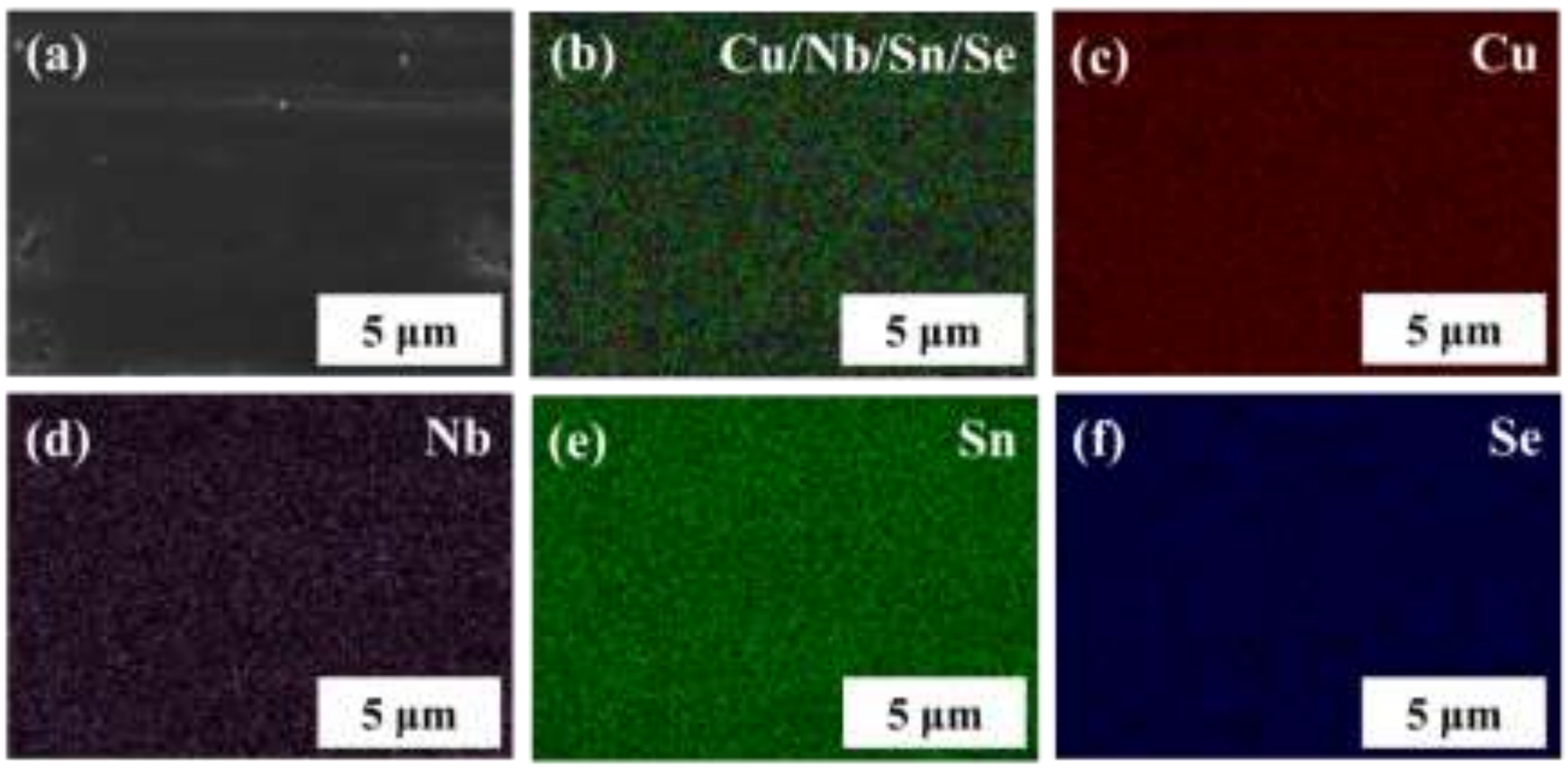

Figure S2. FE-SEM analyses of CNS-$Se_{32}$ sample, (a) top view of polished surface, (b) overlay elemental mapping, and (c) - (f) the individual contribution of elements.

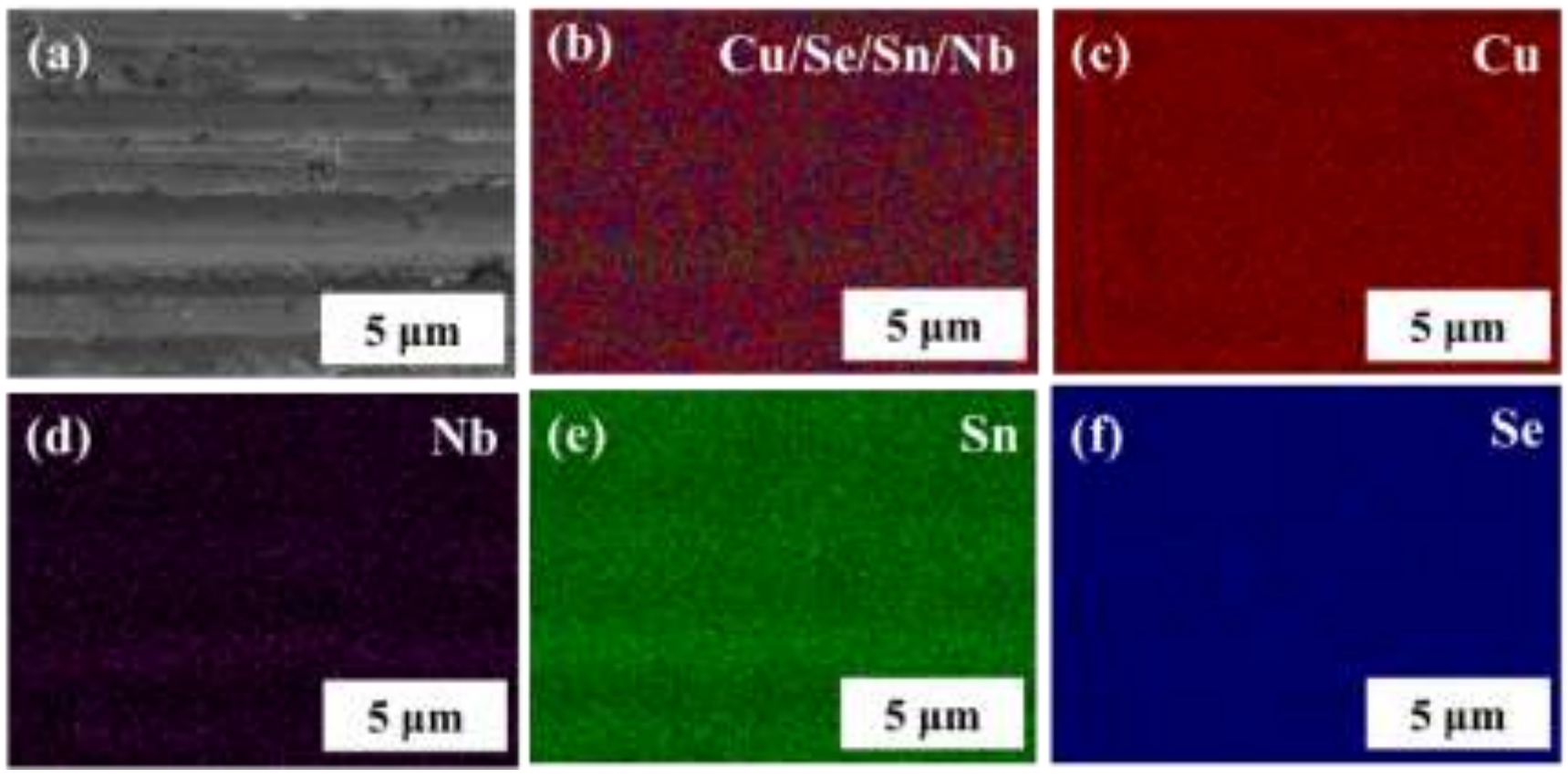

Figure S3. FE-SEM images of CNS-$Se_{31.5}$ sample, (a) top view of polished surface, (b) overlay elemental mapping, and (c) - (f) the individual contribution of elements.

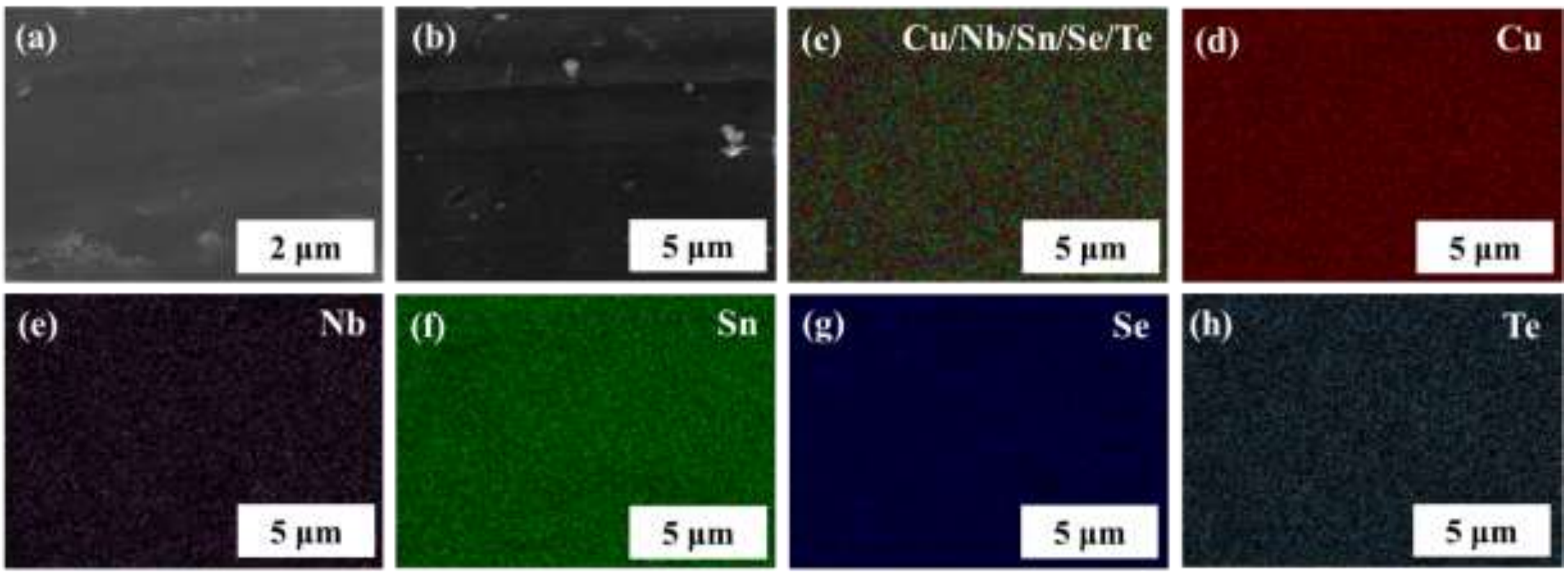

Figure S4. FE-SEM images of (a) CNS-$Se_{30}Te_2$ sample, (b) top view of polished surface, (c) overlay elemental mapping, and (d) - (h) the individual contribution of elements.

## Section S4: Representation of $Cu_I$, $Cu_{II}$, $Cu_{III}$, $Se_I$ and $Se_{II}$ atoms in CNS-$Se_{32}$

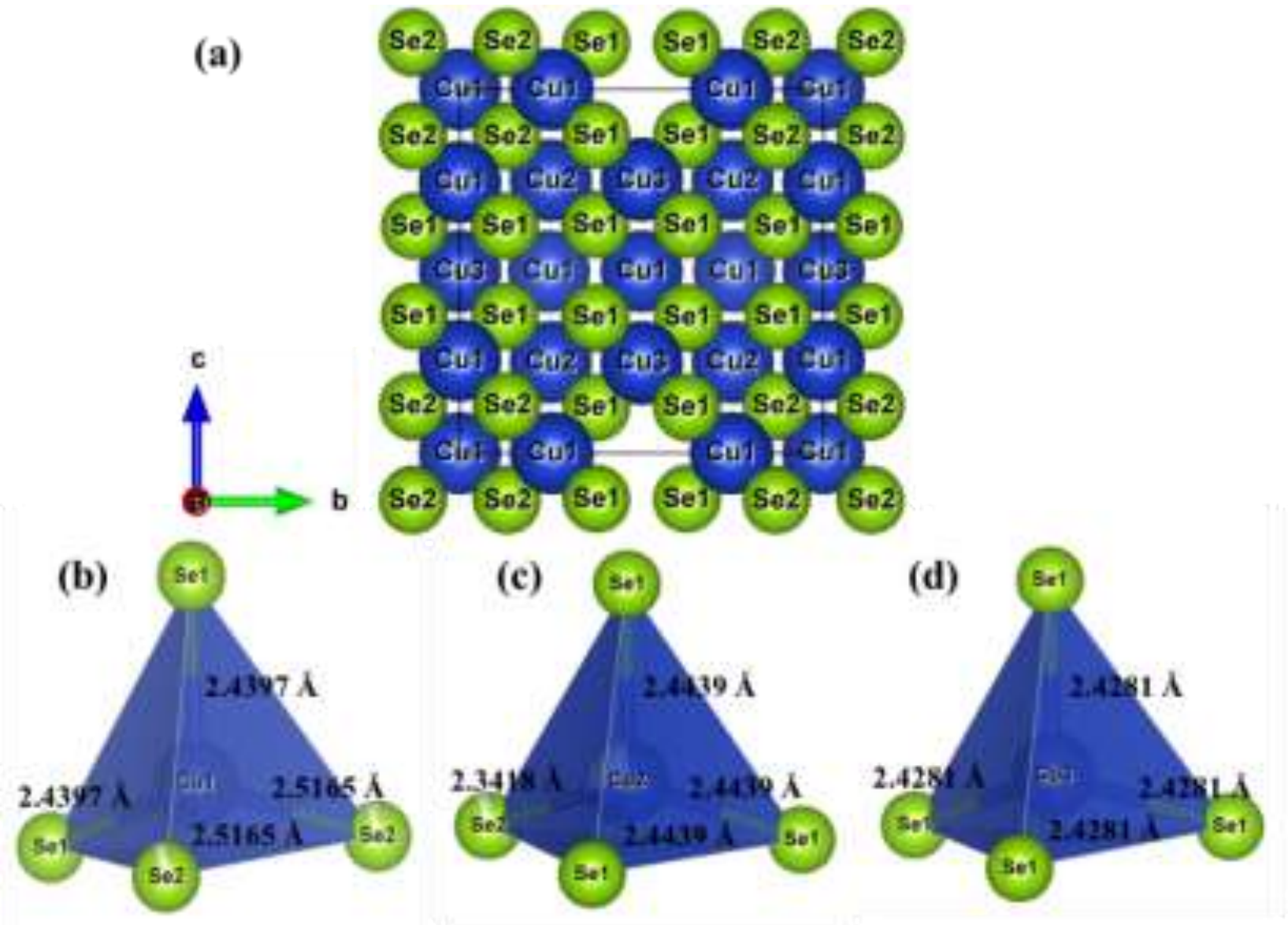

Figure S5. (a) Crystal structure representation of $Cu_I$, $Cu_{II}$ and $Cu_{III}$ with $Se_I$ and $Se_{II}$ atoms in CNS-$Se_{32}$, perpendicular to a-axis, (b) Cu ($Cu_I$, $Cu_{II}$ and $Cu_{III}$) and Se ($Se_I$ and $Se_{II}$) tetrahedra with different bond lengths.

**Table S2:** Bond distance between $Cu_I$, $Cu_{II}$ and $Cu_{III}$ with $Se_I$ and $Se_{II}$ atoms in CNS-$Se_{32}$.

| Cu-Se bond | Bond distance (Å) |
|---|---|
| **$Cu_I$-$Se_I$** | 2.43971(7) |
| **$Cu_I$-$Se_{II}$** | 2.51655(7) |
| **$Cu_{II}$-$Se_I$** | 2.44395(7) |
| **$Cu_{II}$-$Se_{II}$** | 2.34182(7) |
| **$Cu_{III}$-$Se_I$** | 2.42810(7) |

## Section S5: Raman spectroscopy analysis

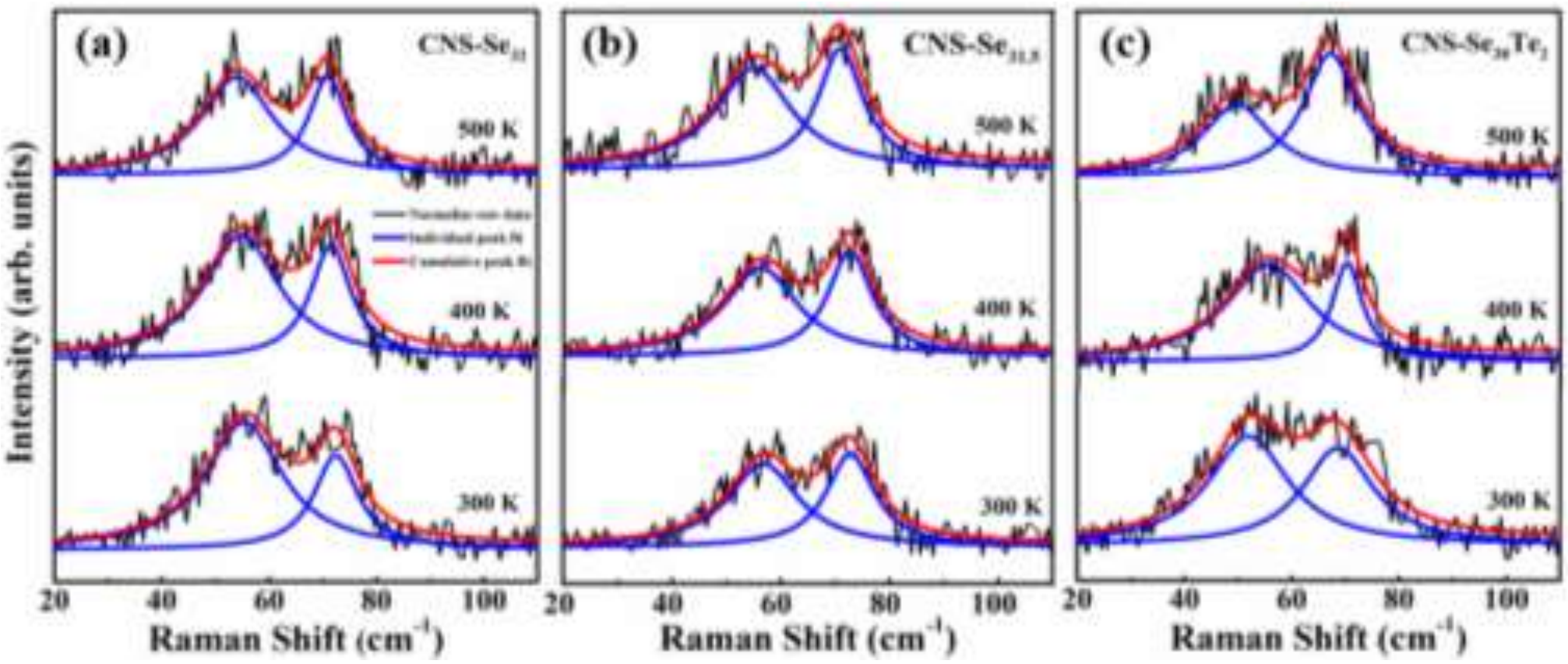

Figure S6. Temperature dependent Raman spectra of (a) CNS-$Se_{32}$, (b) CNS-$Se_{31.5}$ and (c) CNS-$Se_{30}Te_2$.

**Table S3**: Phonon-lifetime ($\tau_i$) of CNS-$Se_{32}$.

| **Phonon Lifetime (in *ps*)** | | | |
|---|---|---|---|
| **Active modes** | **300 K** | **400 K** | **500 K** |
| **$T_1$ (~ 55 $cm^{-1}$)** | 0.31 | 0.31 | 0.33 |
| **$T_1$ (~ 73 $cm^{-1}$)** | 0.53 | 0.58 | 0.61 |

## Section S6: Total density of states

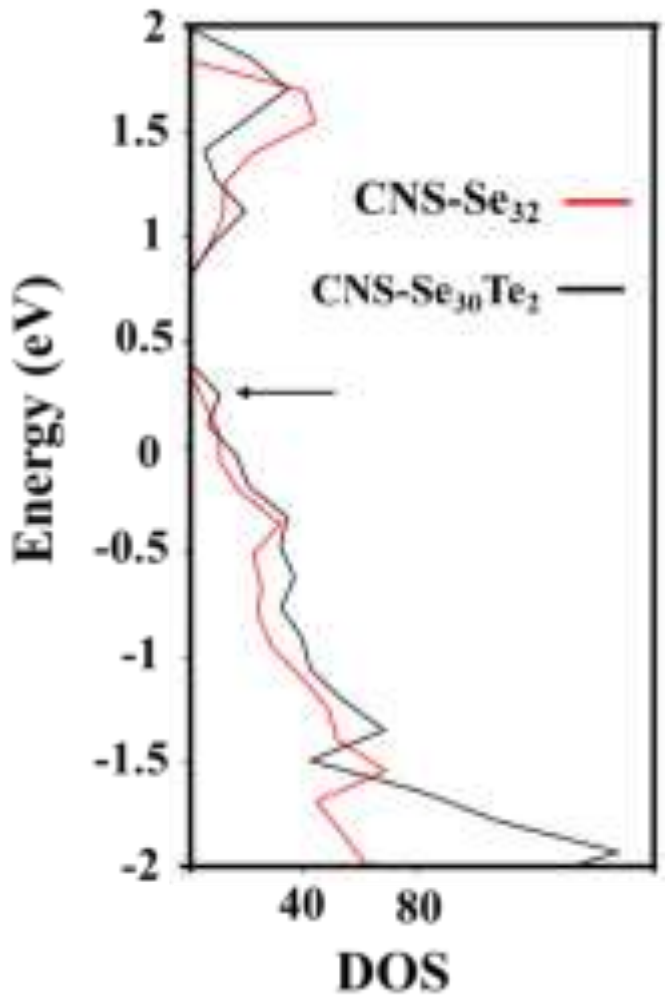

Figure S7. Total density of states of CNS-$Se_{32}$ and CNS-$Se_{30}Te_2$.

## Section S7: Thermal transport analysis

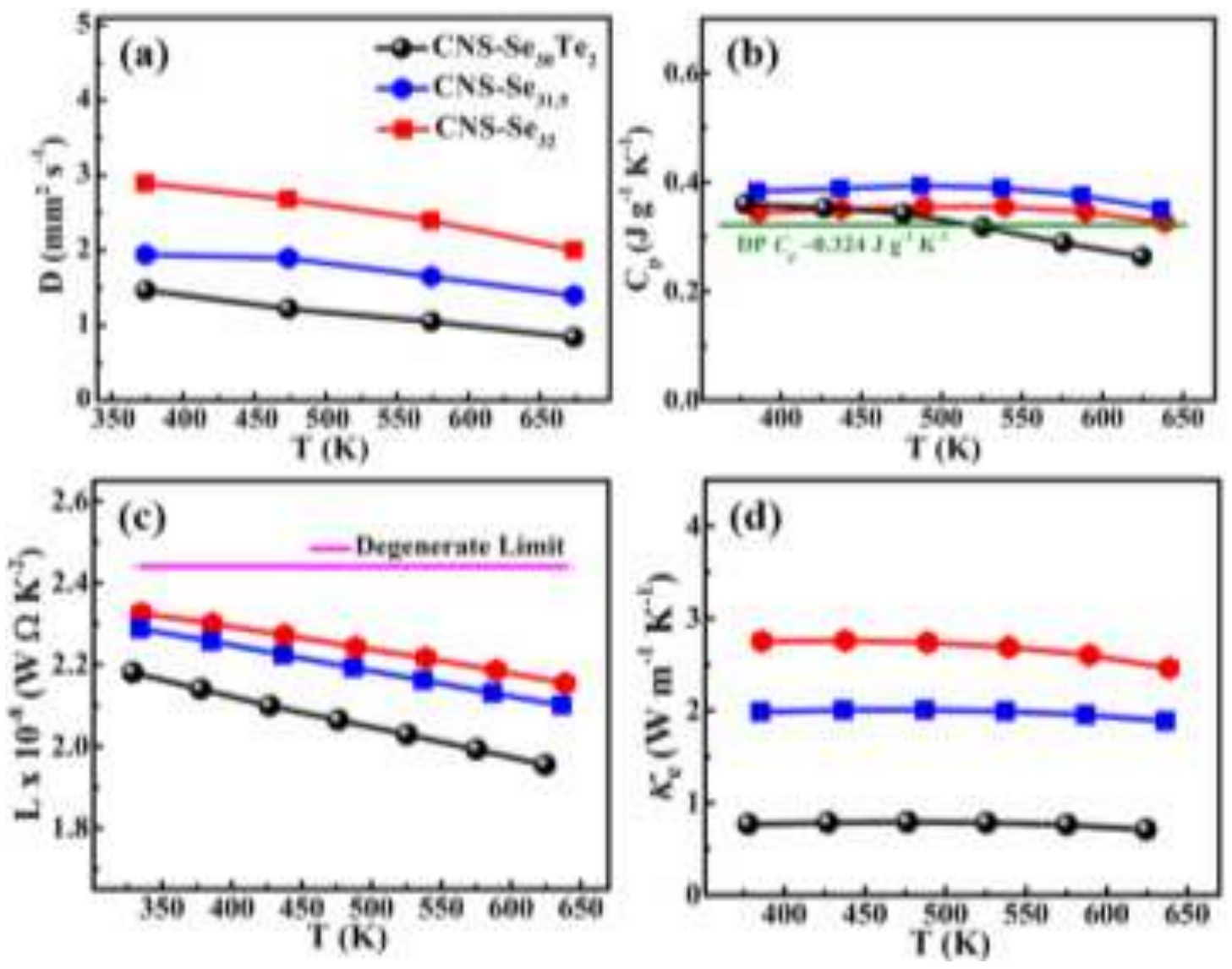

Figure S8. (a) Temperature dependence of thermal diffusivity (*D*), (b) specific heat ($C_p$), (c) Lorenz number (*L*) and (d) electronic thermal conductivity ($\kappa_e$).

**Section S8: Heat capacity analysis:** The Debye temperature ($\Theta_D$) is obtained from the relation $\Theta_D = \left(\frac{12\pi^4 pR}{5\beta}\right)^{1/3}$ where *p* be the total number of atoms present per chemical formula unit and *R* be the molar gas constant.[7] The average sound velocity ($\upsilon_{avg}$) is estimated using the relation; $\upsilon_{avg} = \frac{2\pi K_B \Theta_D}{(6\pi^2 n)^{1/3} h}$ where $K_B$, *n* and *h* represents the Boltzmann constant, total number of atoms present per unit volume (number density of atoms) and Planck's constant, respectively.[7]

**Table S4**. Fitting parameters used to fit the heat capacity data with 1*D*-3*E* model for CNS-$Se_{32}$, CNS-$Se_{31.5}$ and CNS-$Se_{30}Te_2$ compound.

| **Fitting parameters (One Debye + Three Einstein)** | **CNS-$Se_{32}$** | **CNS-$Se_{31.5}$** | **CNS-$Se_{30}Te_2$** |
|---|---|---|---|
| **γ (J mol$^{-1}$ K$^{-2}$)** | 0.2489 ± 0.0155 | 0.1271 ± 0.0090 | 0.2037 ± 0.0066 |
| **β (J mol$^{-1}$ K$^{-4}$)** | 0.00462 ± 0.00023 | 0.00321 ± 0.00012 | 0.00356 ± 0.00010 |
| **A (J mol$^{-1}$ K$^{-1}$)** | 28.52 ± 1.75 | 11.40 ± 0.99 | 13.19 ± 0.74 |
| **$\Theta_{E1}$ (K)** | 22.76 ± 0.47 K | 22.54 ± 0.67 K | 22.53 ± 0.42 K |
| **B (J mol$^{-1}$ K$^{-1}$)** | 247.71 ± 15.09 | 150.40 ± 11.08 | 166.99 ± 8.42 |
| **$\Theta_{E2}$ (K)** | 53.68 ± 1.09 K | 53.92 ± 1.19 K | 53.31 ± 0.79 K |
| **C (J mol$^{-1}$ K$^{-1}$)** | 566.15 ± 12.23 | 401.54 ± 6.46 | 409.70 ± 5.12 |
| **$\Theta_{E3}$ (K)** | 97.92 ± 1.98 K | 93.75 ± 1.61 K | 93.03 ± 1.24 K |
| **~ $\Theta_D$ (K)** | 303 K | 342 K | 330 K |
| **$R^2$ (Adj. R-Square)** | 0.99999 | 0.99999 | 0.99999 |
| **$\chi^2$ (Reduced Chi-Square)** | 3.68726E-4 | 1.22446E-4 | 1.08714E-4 |